\documentclass[8pt,preprint]{aastex}

\shortauthors{Bouchet et al.}
\shorttitle
{ \textit{INTEGRAL} Annihilation emission by SPI/INTEGRAL}

\begin{document}

\title{On the morphology of the electron-positron annihilation emission as seen by SPI/INTEGRAL
\thanks{Based on observations with INTEGRAL, an ESA project with instruments and science data centre
funded by ESA member states (especially the PI countries: Denmark, France, Germany, Italy, 
Spain, and Switzerland), Czech Republic and Poland with participation of Russia and USA.} }

\author{L. Bouchet, J.P. Roques, E.Jourdain}
\affil{CESR--UPS/CNRS, 9 Avenue du Colonel Roche, 31028 Toulouse Cedex~04, France}
%\author{\it Received \today ; accepted  \today}

%\altaffiltext{2}{{\it To whom proofs and offprint requests should be sent}
%\email{bouchet@cesr.fr}}

%%%%%%%%%%%%%%%%%%%%%%%%%%%%%%%%%%%%%%%%%%%%%%%%%%%%%%%%%%%%%%%%%%%%%%%%%
\begin{abstract}
The 511 keV positron annihilation emission remains a mysterious component of the high energy emission of our Galaxy.
Its study was one of the key scientific objective of the SPI spectrometer on-board the INTEGRAL
satellite. In fact, a  lot of observing time has been dedicated to
the Galactic disk with a particular emphasis on the central region.
A crucial issue in such an analysis concerns 
the  reduction technique used to treat this huge quantity of data, and more particularly the 
background modeling. Our method, after validation through a variety of tests, is based 
on detector pattern determination per $\sim$ 6 month periods, together with a normalisation variable on a few hour
timescale.
The Galactic bulge is detected at a level of 70 $\sigma$ allowing
more detailed investigations. 
The main result is that the bulge  morphology can be modelled with two axisymmetric Gaussians
of $3.2^\circ$ and $11.8^\circ$ FWHM and respective fluxes of 2.5 and 5.4 $\times 10^{-4}~photons~cm^{-2}~s^{-1}$.
We found a possible shift of the bulge centre towards negative longitude at $l=-0.6\pm0.2^\circ$.
In addition to the bulge, a more extended structure  is detected significantly
with flux ranging from 1.7 to 2.9 $\times 10^{-3}~photons~cm^{-2}~s^{-1}$ depending on its assumed 
geometry (pure disk or disk plus halo). The disk emission is also found to be symmetric within the limits of the 
statistical errors.

\end{abstract}      

%%%%%%%%%%%%%%%%%%%%%%%%%%%%%%%%%%%%%%%%%%%%%%%%%%%%%%%%%%%%%%%%%%%%%%%%%

%%%%%%%%%%%%%%%%%%%%%%%%%%%%%%%%%%%%%%%%%%%%%%%%%%%%%%%%%%%%%%%%%%%%%%%%%
\section{Introduction}
%%%%%%%%%%%%%%%%%%%%%%%%%%%%%%%%%%%%%%%%%%%%%%%%%%%%%%%%%%%%%%%%%%%%%%%%%
The detection of the 511 keV positron annihilation line emission from the central region of our Galaxy was one of 
the early and important successes of gamma-ray astronomy (Leventhal et al., 1978). Although positron 
annihilation gives rise to the strongest $\gamma$-ray line signal from our Galaxy, three decades of dedicated 
observational and theoretical efforts failed to unveil the origin of the Galactic positrons.  

With existing instrumentation, the emission appears to be diffuse; no point sources of annihilation radiation 
have yet been detected (Malet et al., 1995; Milne et al., 2000; Kn\"odlseder et al., 2005; Teegarden et al., 2005;
De Cesare et al., 2006;Weidenspointner et al., 2006).
Its spatial distribution seems to be symmetric around the Galactic Centre (bulge), with an extent of
$6-8^\circ$ (Full Width at Half-Maximum) and a 511 keV  flux  
measured to be around (1-3) $\times 10^{-3}~photons~cm^{-2}~s^{-1}$
(Leventhal et al., 1978; Milne et al., 2000; Kinzer et al., 2001; Kn\"odlseder et al., 2005; 
Weidenspointner et al., 2008; Bouchet et al., 2008).

There is no firm conclusion as to the  origin  of the positrons due to the modest angular resolution
and limited sensitivity of $\gamma$-ray instruments.
Further, the theoretical interpretation is complex since 
the distribution of the potential sources of positrons is unknown and many uncertainties remain on the composition of
the interstellar gas, the structure and strength of the magnetic field  in the Galaxy and the physics
of positron diffusion and thermalization.

Annihilation radiation from the disk and/or halo is still more difficult to study because of its lower surface brightness
(Milne et al., 2000; Kinzer et al., 2001) but it potentially provides complementary clues to 
the positron production processes involved. Only a few instruments have been able to spatially 
resolve the disk emission from the brighter bulge one. Previous sparse measurements had
shown that the annihilation emission from the disk is brighter in the longitude range $|l| < 18-35^\circ$
with very poor indications on the latitude extent  (Gehrels et al., 1991;
Milne et al., 2000; Kinzer et al., 2001).  
Using more than four years of data from the SPI/INTEGRAL imaging spectrometer, 
its spatial distribution has been  constrained in the longitude range $|l| < 100^\circ$ and latitude extent 
$|b| < 10^\circ$ (Weidenspointner et al., 2008).

In the following, we first present briefly the instrument and data selection. Then, we describe the
analysis method developed for our purpose. The background determination being  a critical issue for  
high energy data,   particularly for extended emission and when the data encompasses a very long period,
we deeply investigate this point and make extensive statistical tests to validate our results.
Finally, the 511 keV sky distribution is derived using two approaches: an imaging method and a sky model fitting
allowing to test several parameterized model distributions representing the Galactic bulge, disk and/or halo.
All the results are discussed and summarized in the last part.

%%%%%%%%%%%%%%%%%%%%%%%%%%%%%%%%%%%%%%%%%%%%%%%%%%%%%%%%%%%%%%%%%%%%%%%%%
\section{Instrument and observations}
%%%%%%%%%%%%%%%%%%%%%%%%%%%%%%%%%%%%%%%%%%%%%%%%%%%%%%%%%%%%%%%%%%%%%%%%%
The ESA's INTEGRAL observatory was launched from Baikonour, Kazakhstan, on 2002 October 17. The spectrometer 
SPI (Vedrenne et al. 2003) observes the sky in the 20 keV to 8 MeV energy range with an energy resolution 
ranging from 2 to 8 keV. It consists of an array of 19 high-purity germanium detectors operating around 80 K,
the spectroscopic performance being maintained thanks to regular annealings of the detection plane.
Its geometric surface is 508~cm$^2$ with a thickness of 7 cm. The assembly is surrounded by a 5 cm thick bismuth
germanate (BGO) which provides background rejection. In addition to its spectroscopic capability, 
SPI can image the sky with a spatial resolution of $\sim 2.6^\circ$  (FWHM) over a field of view of $30^\circ$, 
thanks to a coded mask located 1.7 m above the detector plane.
The instrument's in-flight performance is described in Roques et al. (2003). 
Because of the small number of detectors, SPI's imaging capability relies on a specific observational 
strategy, which is based on a dithering procedure by steps of $\sim 2^\circ$. Each
pointing lasts between 30 and 60 minutes. We have analyzed observations recorded from 2003, February 22 to 2009 January 2,
covering the entire sky. Data polluted by solar flares and radiation belt entries are excluded.
In order to completely suppress the effects due to radiation belts and to secure 
the data set, we suppress also systematically the first and  last 10 exposures of each revolution.

\begin{figure}%1
\plotone{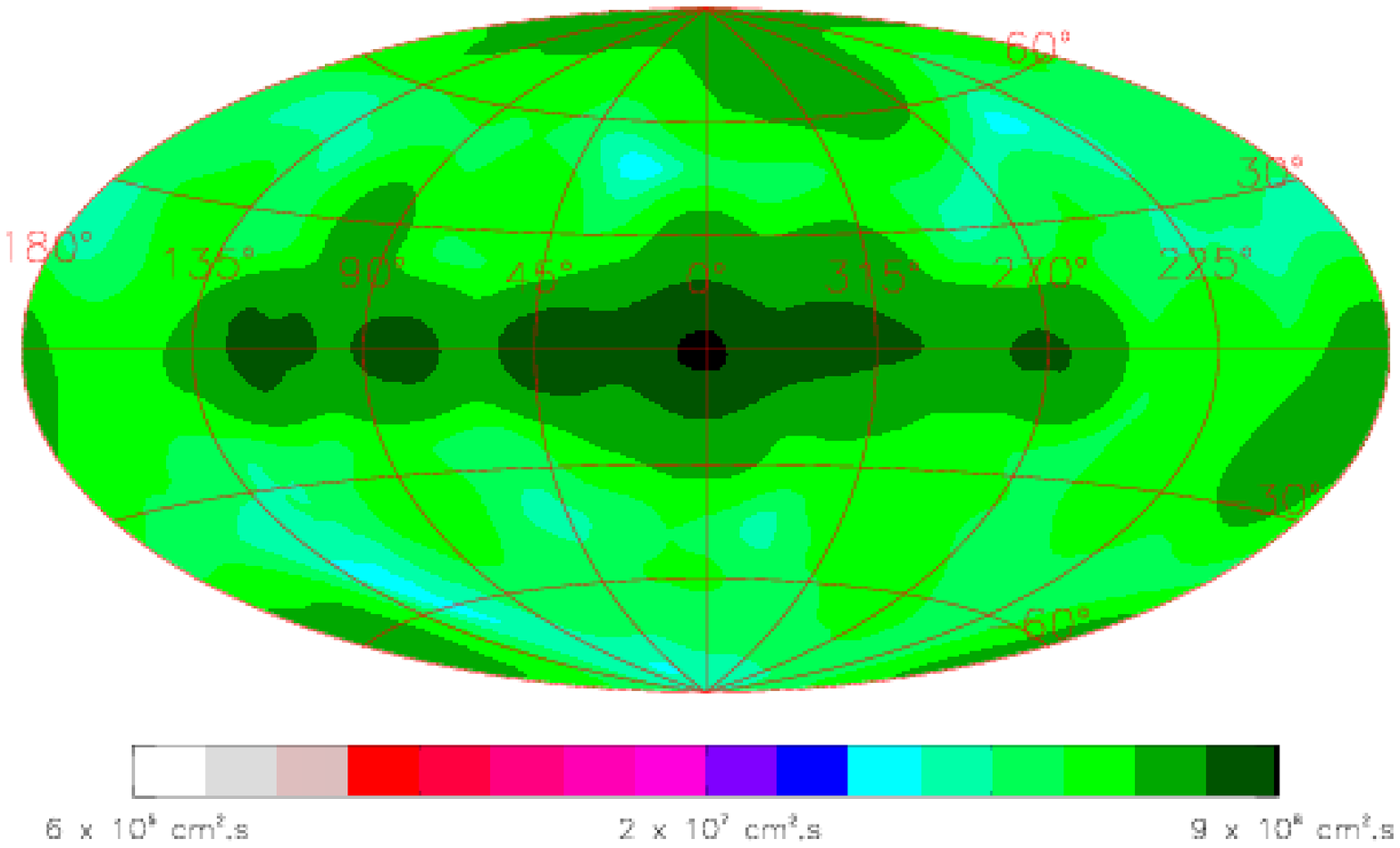}
\caption{508.25-513.75 keV \textit{INTEGRAL} SPI exposure map. Units are in 
$cm^2\times s$. 
This map takes into account the differential sensitivity of SPI across its field 
of view.}
\label{fig:expomap}
\end{figure}

Finally, after image analysis and cleaning, $\sim 1.1 \times 10^{9}~s $ corresponding to 39294 pointings (or exposures)
are kept for the present study. The resulting exposure map is depicted in figure~\ref{fig:expomap}.

%%%%%%%%%%%%%%%%%%%%%%%%%%%%%%%%%%%%%%%%%%%%%%%%%%%%%%%%%%%%%%%%%%%%%%%%%
\section{Data analysis}
%%%%%%%%%%%%%%%%%%%%%%%%%%%%%%%%%%%%%%%%%%%%%%%%%%%%%%%%%%%%%%%%%%%%%%%%%
The signal recorded by the SPI camera on the 19 Ge detectors is composed of contributions from each source (point-like or
extended) in the field of view convolved by the instrument aperture, plus the background.
For extended/diffuse sources, we assume a spatial morphology given by an analytical function or an emission map. 
For $N_s$ sources located in the field of view, the data $D_{p}$ obtained during an
exposure (pointing) p, for a given energy band, can be expressed by the relation:

\begin{equation}
D_{p}=\sum_{j=1}^{N_s} R_{p,j} S_{p,j} + B_{p}
\end{equation}
where $R_{p,j}$ is the response of the instrument for the source j, $S_{p,j}$ is the flux of the source j, 
and $B_{p}$ is the background recorded during the pointing p. $D_{p}$, $R_{p,j}$, 
and $B_{p}$ are vectors of 19 elements. For a given set of $N_p$ exposures, we have to solve a system of $N_p$
equations (1). For that, it is mandatory to reduce the number of unknowns.  
 
A first way to do that takes advantage of the relative stability of the background pattern to rewrite
the background term as: 

\begin{equation}
B_{p} = A_p \times U\times t_{p}
\end{equation}
where $A_p$ is a normalization coefficient per pointing, U is the "uniformity map"
or background count rate pattern on the SPI camera  and 
$t_{p}$ the effective observation time for pointing p. U and t are vectors of 19 elements (one per detector).

The number of unknowns (free parameters) in the set of $N_p$ equations is then $N_p \times (N_s + 1)$ (for the $N_s$ sources and
the background intensities).
Fortunately, a further reduction can be obtained since, at 511 keV, point sources are weak and constant within their error bars.
Concerning the background intensity we consider it variable and test several   timescales.
Finally, we have $N_s + N_b$ ($N_b$ being the number of time bins for the background) free parameters to be determined. 
Practically,  $N_b \backsimeq$ 39000 if the background intensity is allowed to vary on 
the one exposure  timescale  ($\sim$ 2700 s) and $N_b \backsimeq$ 6000 if the background intensity varies 
on a $\sim$6 hour  timescale.

The global background count rate pattern on the SPI camera can also vary but on a much longer timescale (several months).  
We did not find any noticeable pattern variation over 6 months of data, corresponding to the interval between 2 annealings.
With this compromise: 
hours variability for the intensity parameter and modest ($\sim$ 6 months) variability
for the background pattern , we obtain adequate fits to our measurements with an optimized number of parameters.

We performed our analyses in the 508.25-513.75 keV band. Such a 5.5 keV wide band centered at 511 keV 
takes into account the Germanium energy resolution (FWHM $\sim$ 2.05 keV ) including its
degradation  between two consecutive annealings ($5\%$).
At this energy, the gain calibration (performed orbit-wise) accuracy is better than
$\pm$0.01 keV. Finally, only single-detector event data are used (Roques et al., 2003).

%%%%%%%%%%%%%%%%%%%%%%%%%%%%%%%%%%%%%%%%%%%%%%%%%%%%%%%%%%%%%%%%%%%%%%%%%
\subsection{Core algorithm}
%%%%%%%%%%%%%%%%%%%%%%%%%%%%%%%%%%%%%%%%%%%%%%%%%%%%%%%%%%%%%%%%%%%%%%%%%
The tools used for background modeling, imaging, and model fitting were specifically developed 
for the analysis of SPI data and described in Bouchet et al. (2008).
To determine the sky model parameters we adjust the data through a multi-component fitting algorithm, based on the maximum likelihood test statistics.
We used Poissonian statistics to evaluate the quality of various sky models. The core algorithm to
handle such a large, but sparse system is based on MUMPS \footnote{developed by the IRIT/ENSEEIHT laboratory http://mumps.enseeiht.fr}
software together with an error bar computation technique dedicated and optimized for the INTEGRAL/SPI
response matrix structure (Amestoy et al., 2006; Rouet, 2009; Tzvetomila, 2009).

%%%%%%%%%%%%%%%%%%%%%%%%%%%%%%%%%%%%%%%%%%%%%%%%%%%%%%%%%%%%%%%%%%%%%%%%%
\subsection{Background}
%%%%%%%%%%%%%%%%%%%%%%%%%%%%%%%%%%%%%%%%%%%%%%%%%%%%%%%%%%%%%%%%%%%%%%%%%

%%%%%%%%%%%%%%%%%%%%%%%%%%%%%%%%%%%%%%%%%%%%%%%%%%%%%%%%%%%%%%%%%%%%%%%%%
\subsubsection{Modelling}
%%%%%%%%%%%%%%%%%%%%%%%%%%%%%%%%%%%%%%%%%%%%%%%%%%%%%%%%%%%%%%%%%%%%%%%%%
The modeling of the instrumental background is an important issue within the data analysis. 
Since 2 detectors failed (detector 2 on Dec. 7, 2003 and detector 17 on Jul. 17, 2004) during the period spanned by the
observations, the distribution of the instrumental background in the detector plane changes significantly.

This uniformity map or background pattern (eqn. 2) can be derived from empty field observations.
The dedicated INTEGRAL/SPI "empty field" observations are rare. But by making the hypothesis that the 511 keV emission is
essentially concentrated along the Galactic plane and in the bulge, many exposures can be 
considered as empty-field observation at this energy. In practice we built a set of empty-fields using exposures whose
pointing latitude satisfies $|b| > 30^\circ$. 
Furthermore, we exclude pointings closer than  $30^\circ$ from Cyg X-1 and the Crab Nebula since they are potential
emitters in the 505-515 keV energy range.
Finally, 5729 from the 39294 exposures can be used to fix the background patterns.
Table 1 displays periods
between consecutive annealings ($\sim$ 6 months), the revolutions spanned, the total number of exposures,
the number of exposures used to build/test the "empty-field" and the statistics about the location in longitude and
latitude of these exposures.

%%%%%%%%%%%%%%%%%%%%%%%%%%%%%%%%%%%%%%%%%%%%%%%%%%%%%%%%%%%%%%%%%%%%%%%%%
\subsubsection{Basic characteristics}
%%%%%%%%%%%%%%%%%%%%%%%%%%%%%%%%%%%%%%%%%%%%%%%%%%%%%%%%%%%%%%%%%%%%%%%%%
We define the quantity $u$ to quantify the properties of the background pattern for each "empty-field" period,  
 
\begin{equation}
u=\frac {U}{ U_{moy}}~~with~U_{moy}=\frac {1} {N^{eff}}\sum_{d=1}^{N^{eff}} {U(d)} 
\end{equation}
$N^{eff}$ is the number of working detectors (19,18 or 17), u and U  are vectors of $N^{eff}$ elements. u represents
the detector plane count rates normalized to the mean.

\begin{figure}%2
\plotone{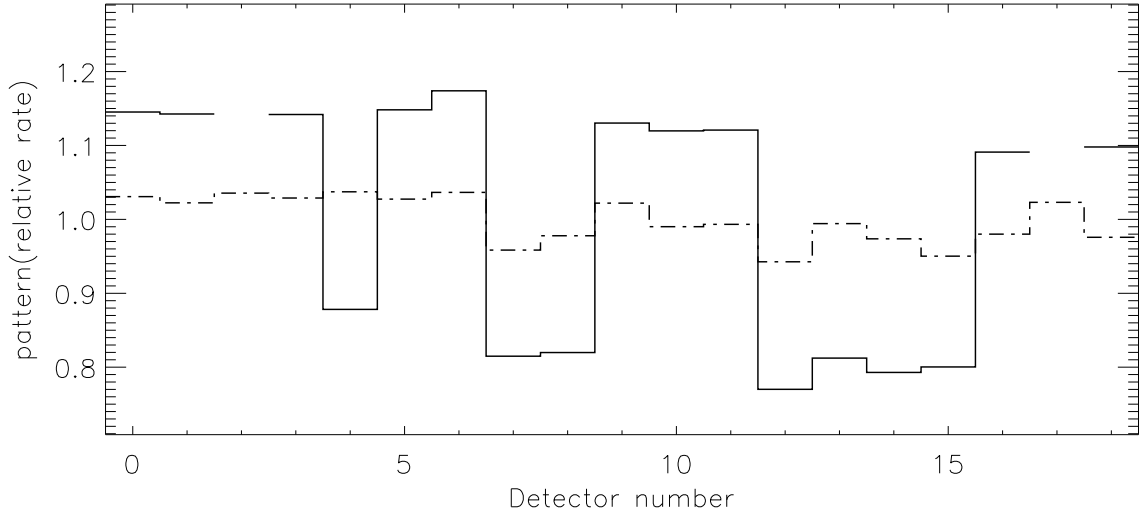}
\caption{508.25-513.75 keV relative pattern for the empty-field corresponding to revolutions 97-131 (dash-dot line) 
and  647-713 (solid line).}
\label{fig:fig5_6588}
\end{figure}

Then, for each empty field period, we determine the background pattern as the time averaged vector u.
Figure~\ref{fig:fig5_6588} displays the resulting pattern for two periods. The difference between the two patterns is
essentially due to the failure of two detectors (2 and 17).

\begin{figure}%3
\plotone{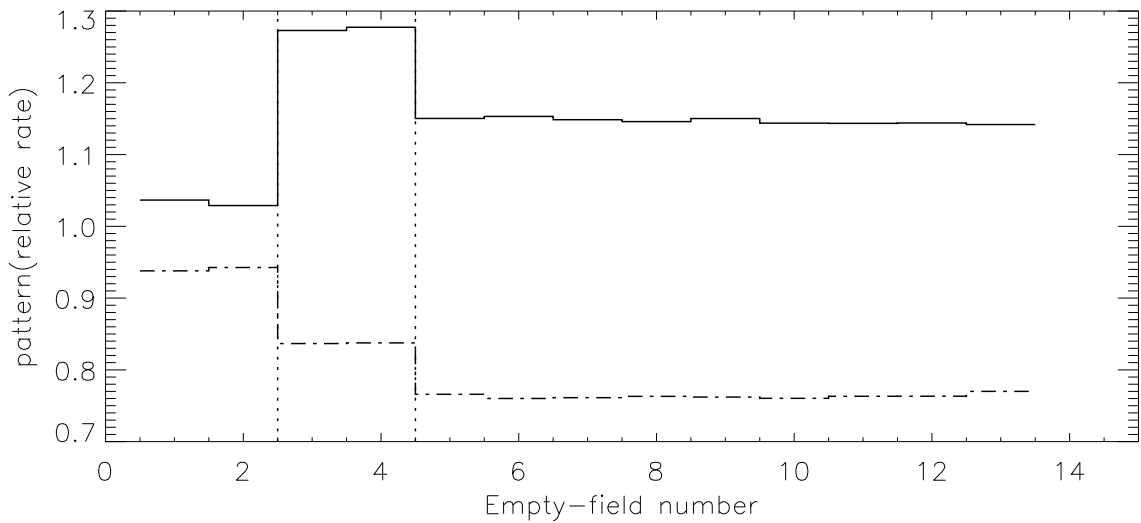}
\caption{508.25-513.75 keV pattern for detector 3 (dash-dot line) and 
12 (solid line) as a function of the empty-field number. The dotted lines indicate 
when detector 2 and 17 failures occured.}
\label{fig:fig6_6588}
\end{figure}

To follow the pattern variation during the mission, we studied the evolution of this quantity for each detector.
Figure~\ref{fig:fig6_6588} displays the evolution of the count rate measured on 2 detectors for each period using the empty field data set. This variation is very small, except at the detector failures.
We also tried to improve the pattern determination by allowing a time dependency or 
by subdividing each "empty field" period in  $\sim$ 3 and $\sim$ 1.5 month period but the F-test   
shows that the fit is not improved.
In consequence, we used the $\sim$ 6 month period interval as defined in table 1 to determine the background pattern.

\begin{figure}%4
\plotone{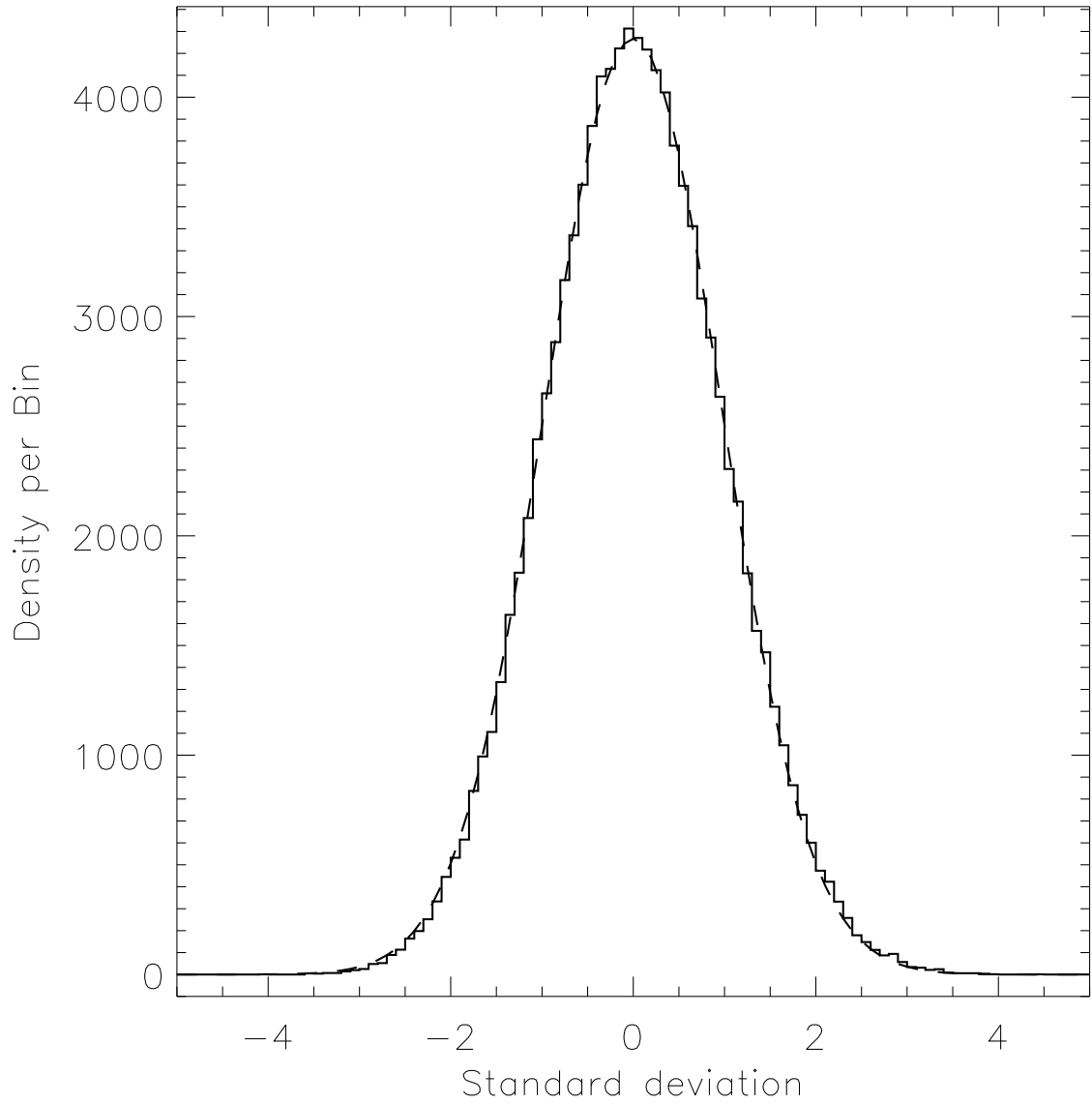}
\caption{508.25-513.75 keV residual distribution for the entire empty field 
dataset (residuals minimum and maximum are respectively -4.02 and 4.46). The 
distribution is compared to a normal distribution with a variance of 0.94. The 
Kolmogorov-Smirnov test gives a distance d=0.005 and a probability p=0.94.
 The run test probability is 0.38}
\label{fig:fig3_511_6588}
\end{figure}

As a final check, the distribution of residuals between our background model and the empty fields data has been tested against various
statistical tests (Kolmogorov Smirnov, run test, moments of the distribution) and exhibit the expected normality properties.
This distribution is plotted on fig.~\ref{fig:fig3_511_6588}, 
compared to a theoretical normal distribution.
This confirms, a posteriori, the hypothesis of a constant pattern between successive annealings and the absence
of any significant source.

Another way to determine the background pattern is to solve eqn 1 for the source and the background simultaneously.
While this method provides a better $\chi^2$ (the global model being more flexible), the resulting patterns are model
dependent. This is understandable in terms of an unavoidable cross-talk between the extended source emission and the
background pattern. As a consequence, the large scale emissions are partially suppressed and the statistical 
differences between the  models are reduced.
Futhermore, the patterns obtained in such way are not perfectly compatible with the  empty
field data set.

%%%%%%%%%%%%%%%%%%%%%%%%%%%%%%%%%%%%%%%%%%%%%%%%%%%%%%%%%%%%%%%%%%%%%%%%%
\subsubsection{Background intensity variability }
%%%%%%%%%%%%%%%%%%%%%%%%%%%%%%%%%%%%%%%%%%%%%%%%%%%%%%%%%%%%%%%%%%%%%%%%%
The backgound model has been tested on our total dataset (39294 exposures, see table 1).
The sky model consists of two point sources, namely
Cyg X-1 and the Crab nebula, two axisymmetric Gaussians of FWHM 3.2$^\circ$ and 11.8$^\circ$ 
to describe the bulge and a Robin disk 1-2 Gyr (Robin et al., 2003) to represent the spatial morphology of the disk. 
Eqn 1 is solved for $S_j$ and $A_p$ , $U$ being fixed and computed from the empty-field dataset
(5729 exposures) as described above.

\begin{figure}%5
\plotone{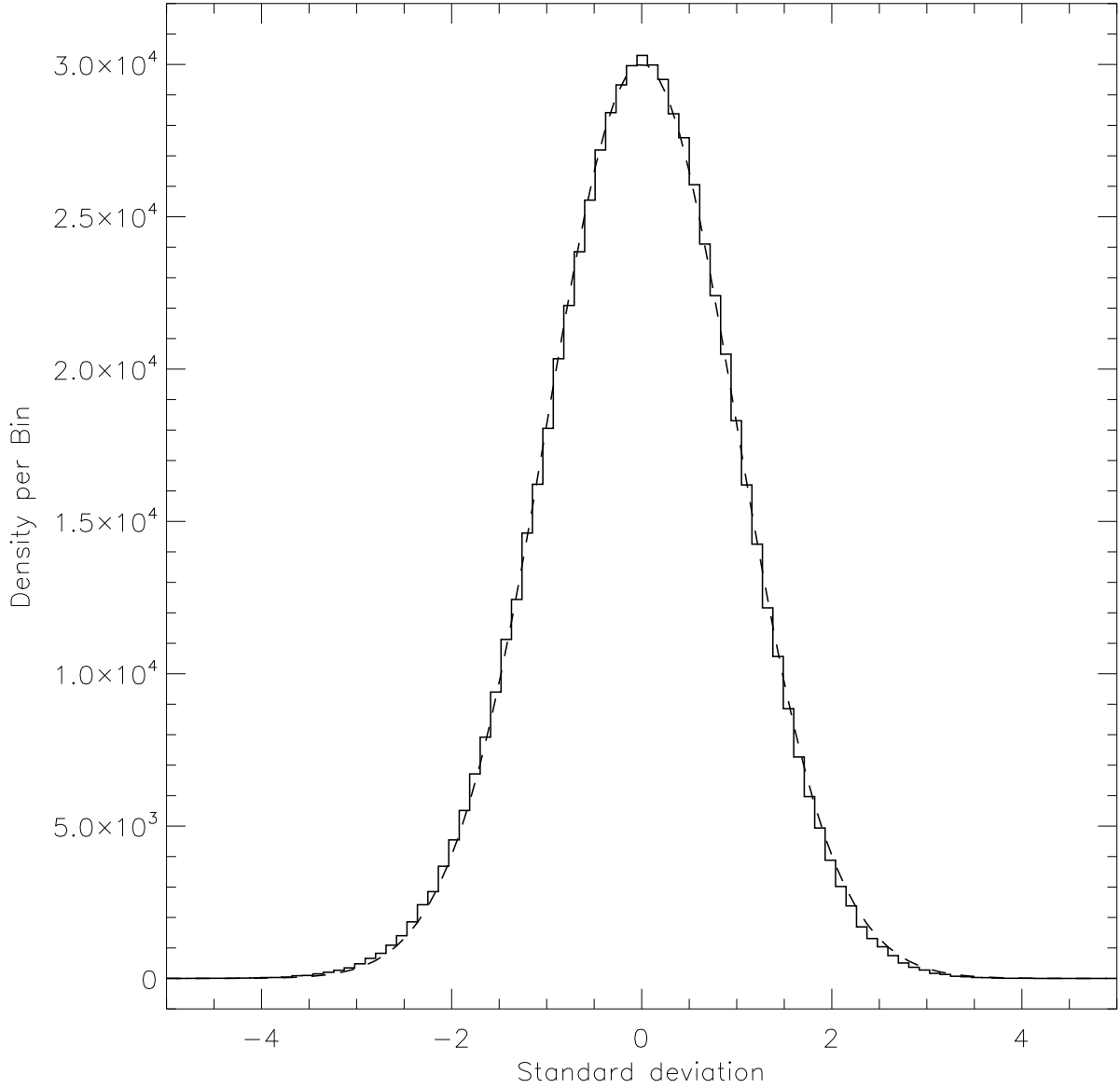}
\caption{508.25-513.75 keV residual distribution for our reference sky model 
(residual minimum and maximum are respectively -4.97 and 4.36. The distribution 
is compared to a normal distribution having a variance of 0.996. The Kolmogorov-
Smirnov test gives a distance d=0.0007 and a probability p=0.999, the run test 
probability is 0.05. }
\label{fig:fig3_511_39294}
\end{figure}

First, the background intensity $A_p$ is assumed to vary on a timescale of $\sim$ 6 hours. The residuals distribution between our model
(sky plus background) and the data appears quite satisfactory (Fig.~\ref{fig:fig3_511_39294}).
Secondly, $A_p$ is allowed to vary on the pointing timescale ($\sim$ 1800-3600 s).
The F-test shows that the second option does not improve the fit. In addition, the $\sim$ 6 hour timescale analysis produces 
slightly reduced error bars (less free parameters) for background and source components while their intensities are 
perfectly compatible with those obtained with a more variable background. 

Finally, the standard configuration used for the
subsequent analysis consists of a fixed bakground pattern per period between two annealings with an intensity variable
on a 6 hours timescale.

%%%%%%%%%%%%%%%%%%%%%%%%%%%%%%%%%%%%%%%%%%%%%%%%%%%%%%%%%%%%%%%%%%%%%%%%%
\subsection{Investigation of the bulge morphology}
%%%%%%%%%%%%%%%%%%%%%%%%%%%%%%%%%%%%%%%%%%%%%%%%%%%%%%%%%%%%%%%%%%%%%%%%%
%%%%%%%%%%%%%%%%%%%%%%%%%%%%%%%%%%%%%%%%%%%%%%%%%%%%%%%%%%%%%%%%%%%%%%%%%
\subsubsection{Maps}
\begin{figure}%6
\plotone{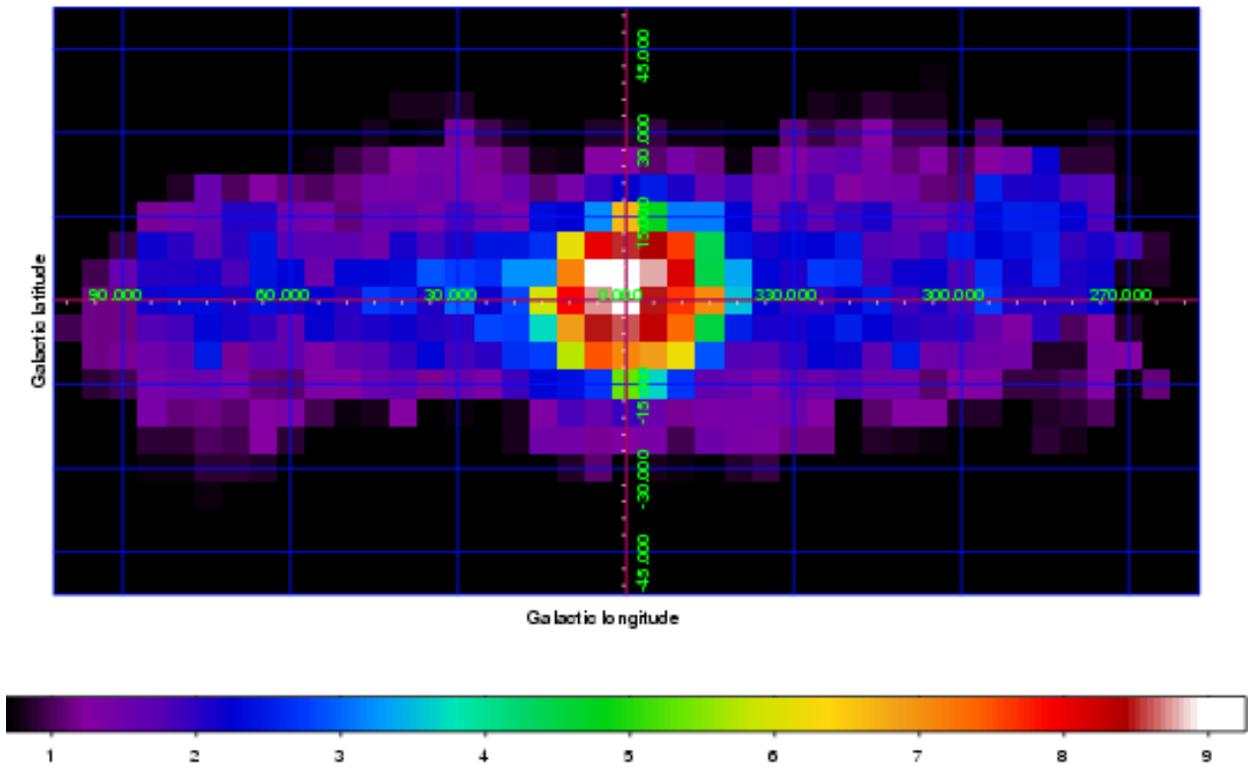}
\caption{508.25-513.75 keV \textit{INTEGRAL} SPI smoothed (top hat of 2 pixels) 
significance map. 
Pixel size is $5^\circ \times 5^\circ$.}
\label{fig:508-514_sigma_map}
\end{figure}

\begin{figure}%7
\plotone{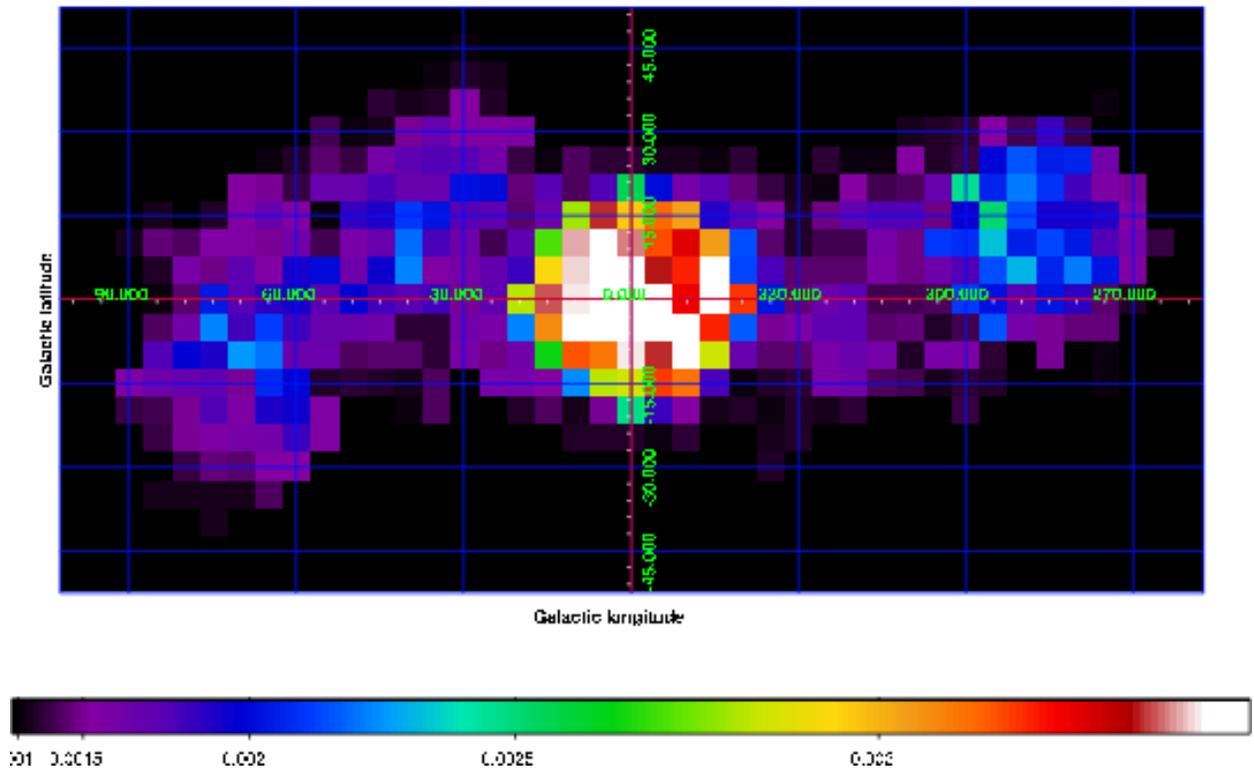}
\caption{508.25-513.75 keV  \textit{INTEGRAL} SPI smoothed (top hat of 2 pixels) 
intensity map in $photons~cm^{-2}~s^{-1}$.
Pixel size is $5^\circ \times 5^\circ$. }
\label{fig:508-514_intensity_map}
\end{figure}

%%%%%%%%%%%%%%%%%%%%%%%%%%%%%%%%%%%%%%%%%%%%%%%%%%%%%%%%%%%%%%%%%%%%%%%%%
The maps provide a model independent view of the sky but each pixel intensity is poorly determined due to the large
number of unknowns. To gather reliable information,
a 508.25-513.75 keV sky map has been built using cells of size  $\delta l = 5^\circ$ x  $\delta b = 5^\circ$ 
(figures~\ref{fig:508-514_sigma_map} and~\ref{fig:508-514_intensity_map}).
 
\begin{figure}%8
\plotone{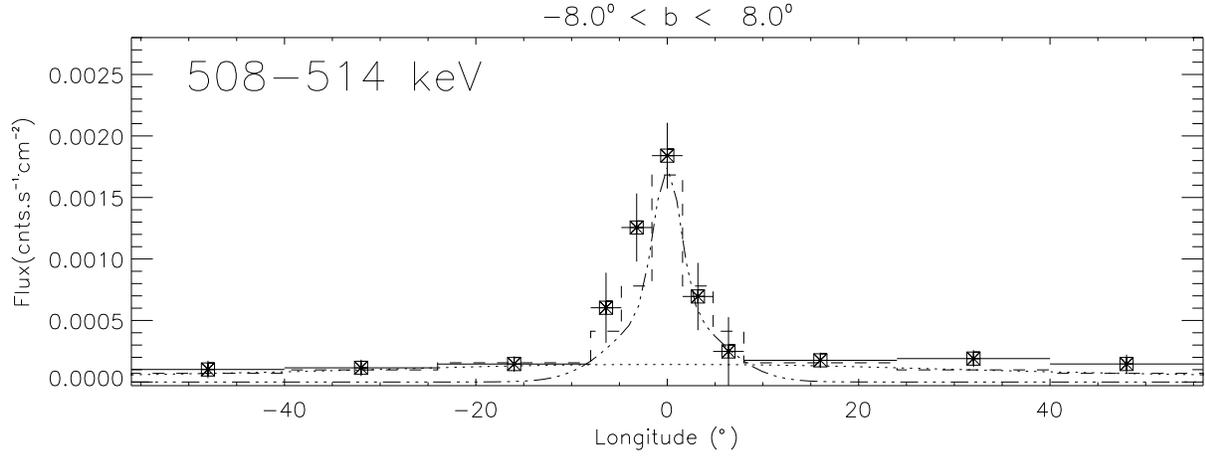}
\caption{508.25-513.75 keV longitude profile for $|b| < 8^\circ$.
Dotted-dashed line corresponds to the sum of  $3.2^\circ$ and $11.8^\circ$ 
axisymmetric Gaussians. 
Dotted line corresponds to the extended distribution (Robin 1-2 Gyr disk model).
The sum of both has been integrated on the same bins as the data (dashed histogram). 
The sky is divided into $16^\circ \times 16^\circ$  pixels except the central 
one which is divided into pixels of $3.2^\circ \times 
3.2^\circ$}
\label{fig:profilong511}
\end{figure}

\begin{figure}%9
\plotone{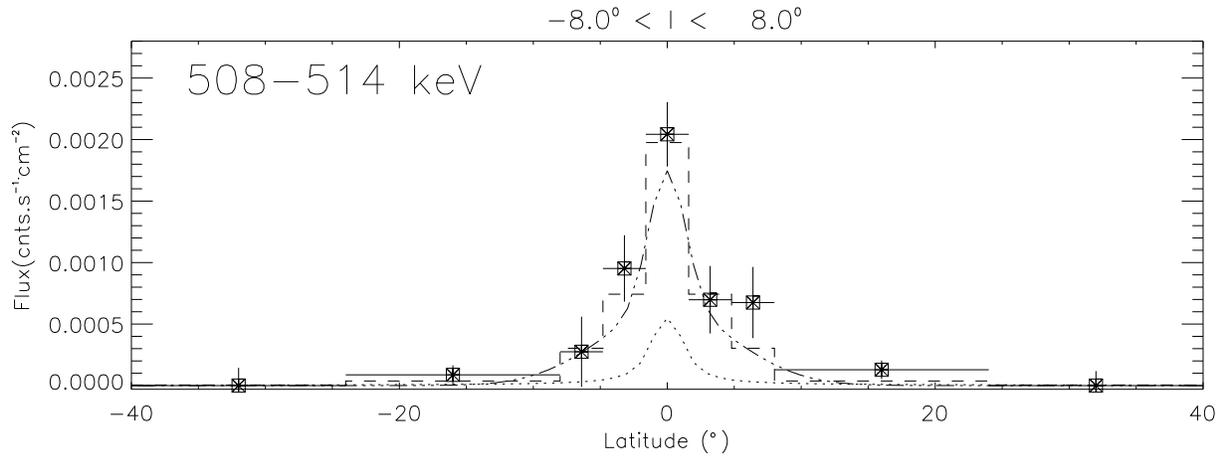}
\caption{Same as fig. 8 but latitude profile for $|l| < 8^\circ$.}
\label{fig:profilat511}
\end{figure}

Emission profiles are also derived using pixels of  $\delta l = 16^\circ$ x  $\delta b = 16^\circ$,
the central one being subdivided into a finer grid (pixel size of
$\delta l = 3.2^\circ$ x  $\delta b = 3.2^\circ$). The reconstruction algorithm described in Bouchet et al. (2008) provides flux and error for each sky pixel.
In this energy range, the emission is almost entirely due to the annihilation line, the contribution of the other 
components being negligible.
The emission profiles (figures~\ref{fig:profilong511} and~\ref{fig:profilat511}) are peaked towards and symmetric around the Galactic Centre. 
However, together with the map, they suggest that the emission is not limited  to a bulge structure but exhibits  
an additional extended component (Galactic disk emission) revealed by the long exposure.

%%%%%%%%%%%%%%%%%%%%%%%%%%%%%%%%%%%%%%%%%%%%%%%%%%%%%%%%%%%%%%%%%%%%%%%%%
\subsubsection{Modelling the bulge with Gaussian profiles}
%%%%%%%%%%%%%%%%%%%%%%%%%%%%%%%%%%%%%%%%%%%%%%%%%%%%%%%%%%%%%%%%%%%%%%%%%
With more than $10^9$ s of exposure time, a single Gaussian (e.g. Kn\"odlseder et al., 2005) does not provide
an acceptable description of the bulge geometry. We obtain a better adjustment of the data with two Gaussians 
of  FWHM 2.6 $^\circ$ and 11.0 $^\circ$, together with an extended component representing the Galactic disk modelled
by a Robin 1-2 Gyr disk ($\sim$220$^\circ$ FWHM longitude extension, $\sim$5.5$^\circ$ FWHM latitude extension).
This two Gaussian model has been used for simplicity.
Adding one or more sources simultaneously to the model, like  GRS1758-258, GS1826-24 or H1743-22 known to emit above 100 keV, 
does not affect significantly the parameters of the Gaussians; none of these sources are detected above $2 \sigma$.
However, a significant change is obtained by introducing 1E1740.7-2942.
It may reflect simply that the bulge profile contains small-scale diffuse emission structures, localized in the central  $\sim3^\circ$.
Furthermore the modest angular resolution of SPI makes it difficult to
distinguish between point sources, point-like and small-scale diffuse emission.
The Gaussian parameters depend weakly on the assumed disk geometry (for example the $240 \mu$ map is slightly peaked 
at the Galactic Center) but remain well inside the error bars.

We also investigate a more complex bulge shape considering different widths in $l$ and $b$. This gives
$\Delta l=(4.0_{-1.0}^{+1.4})$ and $\Delta b=(2.5_{-0.9}^{+1.0})^\circ$  for the smallest Gaussian and
$\Delta l=(9.6_{-2.0}^{+2.5})$ and $\Delta b=(15.2_{-3.0}^{+3.0})^\circ$ for the larger one. The statistics do not
allow the derivation of any meaningful conclusion, but in all cases, these parameters stay compatible with those of the axisymmetric model.

%%%%%%%%%%%%%%%%%%%%%%%%%%%%%%%%%%%%%%%%%%%%%%%%%%%%%%%%%%%%%%%%%%%%%%%%%
\subsubsection{Possible shift of the Gaussian centroid}
\begin{figure}%10
\plotone{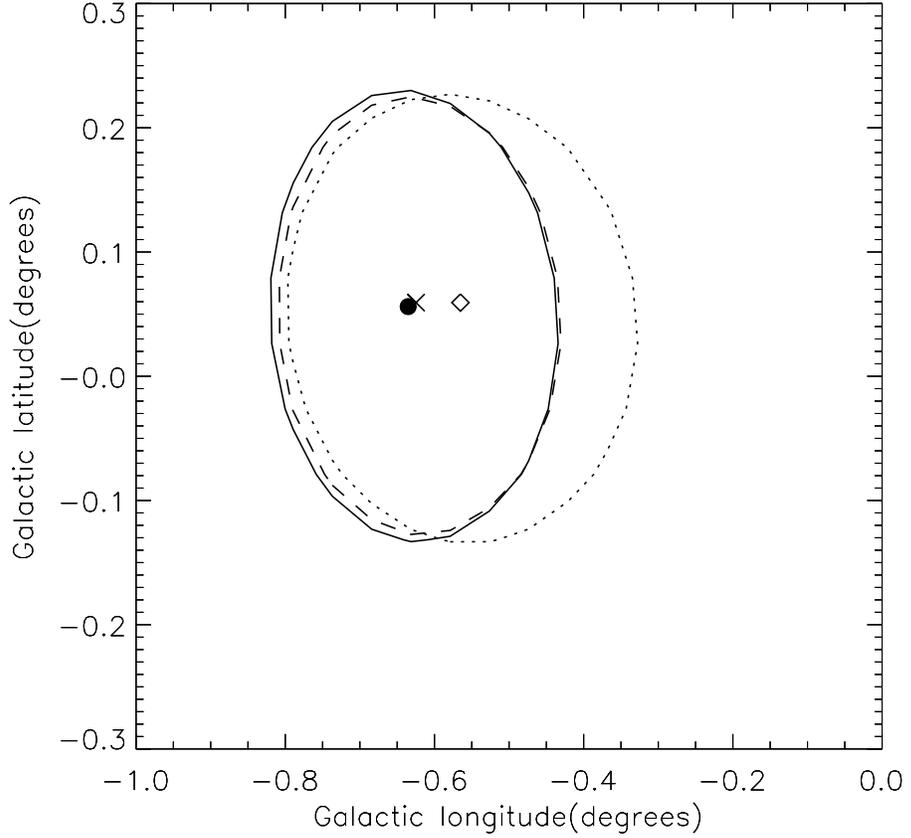}
\caption {2D $1\sigma$ error box on the bulge position for 3 disk models. The bulge is modelled 
with $3.2^\circ$ and $11.8^\circ$ axisymmetric Gaussians. The disk is modelled 
by (a) a Robin 1-2 Gyr disk (solid line),
(b) a Robin 1-2 Gyr disk segmented in 4 parts (short-dashed line) and c) a 
Gaussian with a longitude FWHM of $250^\circ$ and latitude FWHM of  $6^\circ$ 
(long-dashed line)  The best centre is found at (a) $l_0=-0.64 \pm 0.20$ and 
$b_0=0.06 \pm 0.20$, (b) $l_0=-0.57 \pm 0.24$ and $b_0=0.06 \pm 0.20$ and 
(c) $l_0=-0.63 \pm 0.20$ and $b_0=0.06 \pm 0.20$. }
\label{fig:fig_contour}
\end{figure}

%%%%%%%%%%%%%%%%%%%%%%%%%%%%%%%%%%%%%%%%%%%%%%%%%%%%%%%%%%%%%%%%%%%%%%%%%
In a following step, we let free the centroids of the Gaussians $l_0,~b_0$ which are moving together.
The fit gives $l_0=(-0.64_{-0.19}^{+0.21})^{\circ}$ and $b_0=(0.06_{-0.20}^{+0.19})^{\circ}$. 
These values depend slightly on the choosen configuration (in particular we tried several widths of the Gaussians),
but all the parameters remain compatible. Fig.~\ref{fig:fig_contour} shows the 1 $\sigma$ uncertainty
zone in the longitude and the latitute for the centroid. In this configuration the introduction of an additional source
is not needed anymore.

\begin{figure}%11
\plotone{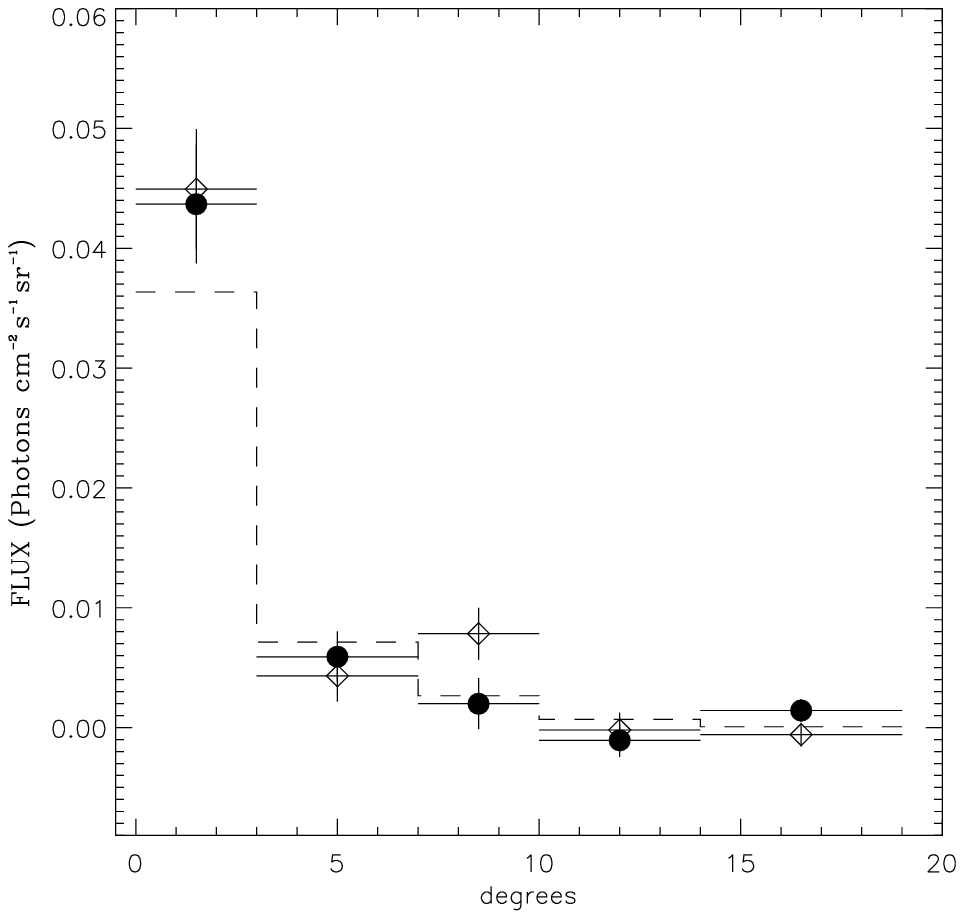}
\caption{508.25-513.75 keV radial profile for $b < 0^\circ$ (filled circles) and 
$b > 0^\circ$ (diamonds).
Dotted-dashed line corresponds to the sum of a $3.2^\circ$ and $11.8^\circ$ 
axisymmetric Gaussians. }
\label{fig:fig_radial_b}
\end{figure}

\begin{figure}%12
\plotone{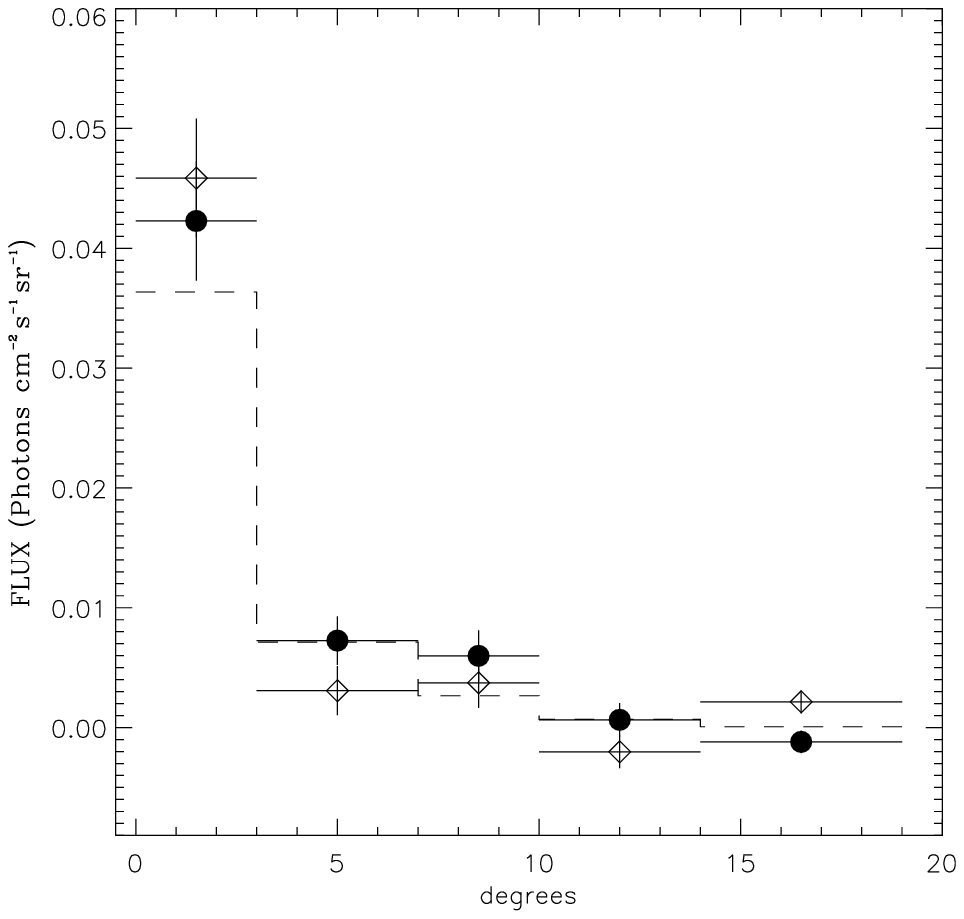}
\caption{508.25-513.75 keV radial profile for $l < -0.6^\circ$ (filled circles) 
and $l > -0.6^\circ$ (diamonds)
Dotted-dashed line corresponds to the sum of a $3.2^\circ$ and $11.8^\circ$ 
axisymmetric Gaussians. }
\label{fig:fig_radial_l}
\end{figure}

To go further, we built a model composed of a set of nested shells of constant density centered at ($l,b$)=(-0.6,0)$^\circ$
with radii 0-3.0, 3-7, 7-10, 10-15 and 15-19$^{\circ}$.  We thus obtain the 511 keV flux distributions just
assuming a radial emission. We split each shell 
into negative and positive part for the longitude and then for the latitude. 
Fig.~\ref{fig:fig_radial_b} and~\ref{fig:fig_radial_l} display the radial profiles we obtained which appear quite compatible 
with the combination of the two off-centered axisymmetric Gaussians determined above.
We thus conclude that the centroid emission appears slightly offset
from the Galactic Centre direction. This result, while marginally significant, 
is all the more remarkable since Kinzer et al. (2001) have found a similar shift of 
the bulge towards negative longitude in the OSSE/CGRO data (1 $\sigma$ shift measured 
between  -0.25 and -1$^{\circ}$).

Finally, we consider that the best configuration (hereafter called the reference model) consists of:
2 Gaussians centered at $l_0=-0.6^\circ$ and $b_0=0.0^\circ$ representing the bulge plus
an extended disk component (Robin 1-2 Gyr).
Then the best fit FWHM of the Gaussians are $(3.2_{-1.0}^{+1.0})^\circ$ and
$(11.8_{-1.5}^{+1.9})^\circ$ with fluxes  of $(2.48 \pm 0.26) \times 10^{-4}~photons~cm^{-2}~s^{-1}$ and
$(5.36 \pm 0.5) \times 10^{-4}~photons~cm^{-2}~s^{-1}$ respectively, the disk flux being
$ (1.83 \pm 0.27) \times 10^{-3}~photons~cm^{-2}~s^{-1}$, providing a 
Bulge to Disk ratio of 0.44. Using the ratio calculated by Kn\"odlseder et al. (2005, Table 3) to infer the 
511 keV line luminosities and the factor ($e^{+}/\gamma_{511})$ = 1.64 from Brown \& Leventhal (1987),
we obtain annihilation rates of $1.1\times 10^{43} s^{-1}$ in the bulge and $0.8\times 10^{43} s^{-1}$
in the disk.

%%%%%%%%%%%%%%%%%%%%%%%%%%%%%%%%%%%%%%%%%%%%%%%%%%%%%%%%%%%%%%%%%%%%%%%%%
\subsection{Disk/halo emission}
\begin{figure}%13
\plotone{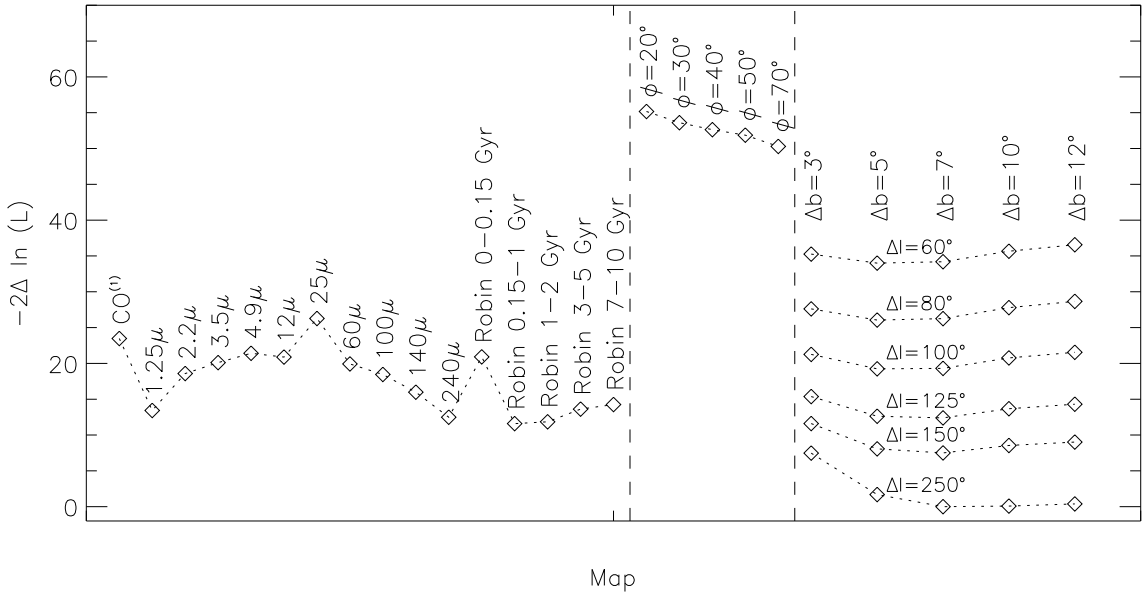}
\caption{Maximum log likelihood-ratio (MLR) as a function of the tracer map.
The left part corresponds to physical maps ($^1$ CO map (Dame et al., 2001)). The central part 
corresponds to pure halo emission modelled with axisymmetric Gaussians. 
In the last part, the disk is assumed to have a Gaussian morphology whose latitude 
and longitude FWHM are indicated horizontally and vertically respectively. 
Gaussians with longitude $>250^\circ$ FWHM are not shown as they do not improve 
the MLR. The bulge is always modelled with the two off-centered Gaussians.}
\label{fig:traceur511}
\end{figure}

%%%%%%%%%%%%%%%%%%%%%%%%%%%%%%%%%%%%%%%%%%%%%%%%%%%%%%%%%%%%%%%%%%%%%%%%%
We then attempted to describe the extended spatial distribution superimposed on the central bulge with various spatial geometries. 
Simple geometric shapes (i.e. two dimensional Gaussians) as well as more physical maps (CO, NIR, Robin disk (Robin et al., 2003)) were tested.\\
Fig.~\ref{fig:traceur511} summarizes the result of the correlation map study, with the $\Delta\chi^2$ ( which varies similarly
 as the reduced maximum log-likelihood ratio) plotted for each of the tracer maps.
Best results are obtained for a Robin disk (in the 0.15-3 Gyr range corresponding to an old stellar population) 
or NIR/DIRBE 240$\mu$ and 1.25$\mu$ maps, which happen 
to be good tracers of the $^{26}Al$ line emission (Kn\"odlseder et al., 1999).
Simple bi-dimensional Gaussians of latitude FWHM $\sim5-7^\circ$ and longitude FWHM $\sim250^\circ$ give 
also good  results. We note that the disk exhibits a larger longitude extension than
the previoulsy reported values: first by OSSE
(Kinzer et al., 2001), but the study is based on a longitudinally
truncated data set distribution and more recently by Weindenspointner et al., 2008. 
Even though it is difficult to describe the emission in greater detail, this result represents a good indication 
for a bulge/disk structure.
Extended disk structure flux (for the most plausible disks) is
around $1.7 \times 10^{-3}~photons~cm^{-2}~s^{-1}$ and the Bulge/disk ratios range from 0.25 to 0.7.
 
This analysis excludes single halo models (modelled by axisymmetric 
Gaussians).
However, in the past, OSSE data have often been  compared with bulge models including some 
non-Gaussian broadening (wings), featuring a bulge +  halo central geometry.
We have tested this kind of configuration by considering a stellar halo (or spheroid) model proposed by Robin et al. (2003).
The central region profile is built following the law:
\begin{equation}
   N~\times~(a/R_{0} )^{m}~     for~ a > a_{c}
\end{equation}   
\begin{equation}
   N~\times~(a_{c}/R_{0})^{m}~for~a < a_{c}
\end{equation}   
\begin{equation}
   a^2=x^2 + y^2  + (z/\epsilon)^2
\end{equation}
N represents a normalisation constant and x,y,z, the cartesian coordinates in the bulge reference frame.      
We obtain a good fit to the data with the axis ratio,~$\epsilon$ = 0.8,~m = -2.6,~$a_{c}$=200~pc
and $R_{0}$=8.5~kpc. The 240$\mu$ and 1.25$\mu$ NIR/DIRBE and Robin 1-3 Gyr maps remain the best tracers of the disk emission.
We note further that assuming these geometries leads to higher B/D ratios (between 1.4 and 2.1 for the best models)
since here B stands for the bulge plus the halo wings.

%%%%%%%%%%%%%%%%%%%%%%%%%%%%%%%%%%%%%%%%%%%%%%%%%%%%%%%%%%%%%%%%%%%%%%%%%
\subsubsection{Disk symmetry}
%%%%%%%%%%%%%%%%%%%%%%%%%%%%%%%%%%%%%%%%%%%%%%%%%%%%%%%%%%%%%%%%%%%%%%%%%
To search for a disk asymmetry, we first used the galactocentric model and we divide the disk (Robin 1-2 Gyr) in 4 parts: 
a) $-180^\circ <$l$< -50^\circ$, b) $-50^\circ <$l$< 0^\circ$, c) $0^\circ <$l$< 50^\circ$ and d) $50^\circ <$l$< 180^\circ$.
The fluxes of these different parts are respectively: (a) ($0.84 \pm 0.27$) (b) ($0.44 \pm 0.07$) (c) ($0.29 \pm 0.07$)
and (d) ($1.31 \pm 0.25$) $10^{-3}~photons~cm^{-2}~s^{-1}$. 
Those values, while compatible with those of Weidenspointner at al. (2008), support a symmetric disk 
within the quoted errors with a west to east ratio of 1.5$\pm$0.4.
In fact, the main difference between both analysis
concerns the background treatment. In Weidenspointner at al. (2008),
the background pattern is determined once per orbit and simultaneously with the sky fluxes. It thus becomes model 
dependent, inducing additional uncertainties in the results (see discussion in section 3.2.2).

Furthermore, we note that the disk flux in $|l| < 50^\circ$ depends on the assumed bulge geometry and position and is hence model-dependent.
Now, using our best model for the bulge, i.e. the off-centered Gaussian described in 3.3.3, the split disk fluxes in region a) b) c) and d) are then ($0.82 \pm 0.27$),
($0.38 \pm 0.07$) ,($0.34 \pm 0.07$) and ($1.32 \pm0.25$) $10^{-3}~photons~cm^{-2}~s^{-1}$ with a west to east ratio of 1.1 $\pm0.4$.

\begin{figure}%14
\plotone{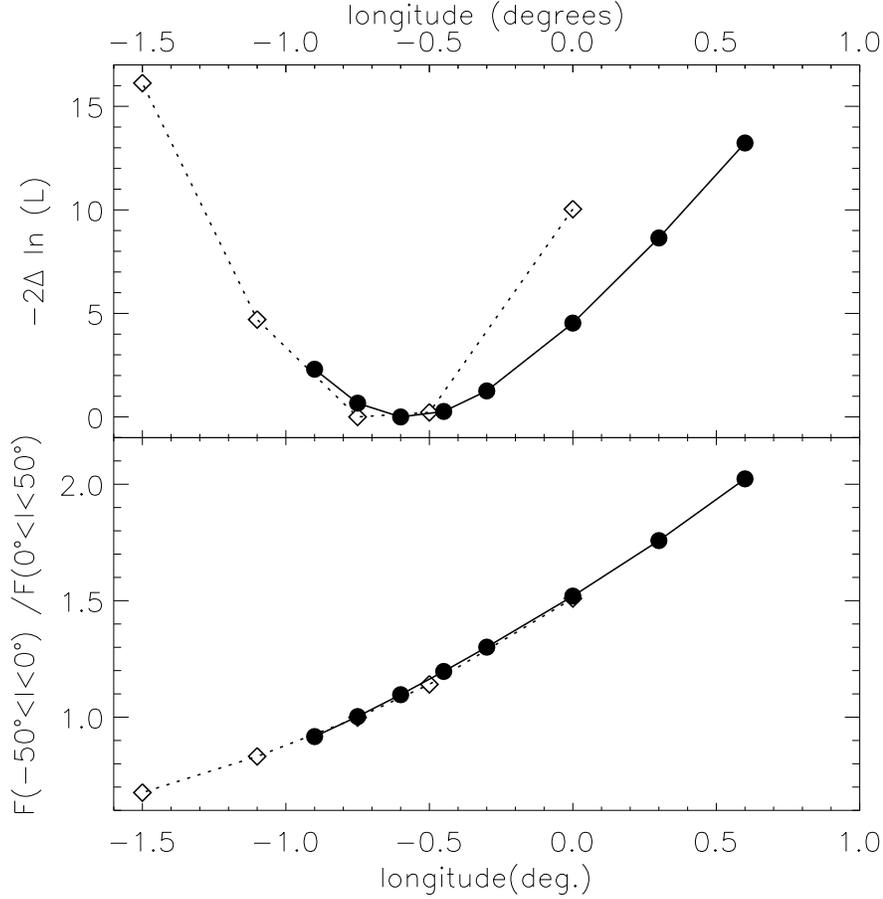}
\caption{Top panel: MLR value as a function of the bulge centre position in longitude.
Bottom: ratio of flux in$-50^\circ <$l$< 0^\circ$ to $0^\circ <$l$< 50^\circ$
as a function of the bulge centre position.
The curves depict the results for two models:
a) The bulge is modelled with two axisymmetric Gaussians of $3.2^\circ$ and 
$11.8^\circ$, the disk with a 1-2 Gyr Robin disk (solid line and filled circle).
b) The longitude and latitude FWHM of the Gaussians modeling the bulge have been fitted for each
longitude position. 
The disk is modelled by a bi-dimensional Gaussian of FWHM 250$^\circ \times 6^\circ$ (Dotted line and diamonds)}
\label{fig:fig_ratio}
\end{figure}

We have investigated the dependency of this ratio as a function of the bulge centroid. 
Fig.~\ref{fig:fig_ratio} displays
the strong correlation between these 2 values and clearly  proves  that an asymmetric disk is not required
by the data.

%%%%%%%%%%%%%%%%%%%%%%%%%%%%%%%%%%%%%%%%%%%%%%%%%%%%%%%%%%%%%%%%%%%%%%%%%
\section{Summary and conclusions}
%%%%%%%%%%%%%%%%%%%%%%%%%%%%%%%%%%%%%%%%%%%%%%%%%%%%%%%%%%%%%%%%%%%%%%%%%
We presented a detailed investigation of the morphology  of the Galactic emission in the narrow annihilation line at 511 keV.
The period covered by the data encompassing more than 6 years, the background determination is a key point in the analysis.
We have chosen to describe the background with a series of fixed patterns (about one per 6 months), together with a 
normalisation adjusted by timebins of 6 hours. The patterns have been measured  in the sky regions outside the Galactic
disk, with the advantage to be independent of the sky model. This method has been validated through various statistical tests 
and represents a major difference between our analysis and that presented in Weidenspointner (2008).
   
While the reality might be more complex, the bulge morphology is  well described by two axisymmetric Gaussians, centered
at $l_0=-0.64$$^\circ$ and $b_0=0.06$$^\circ$, with 
FWHM of 3.2$^\circ$ and 11.8$^\circ$ and fluxes  of $(2.48 \pm 0.26)$  and $(5.36 \pm 0.50) \times 10^{-4}~ 
photons~cm^{-2}~s^{-1}$ respectively.
While marginally significant, it is similar to the shift of the bulge towards negative longitudes 
seen in the OSSE data (Kinzer et al., 2001). 

We confirm also the existence of a disk emission with a longitude extent over $150-250^\circ$ and latitude over $5-7^\circ$.

Many spatial distributions can be used to model the disk/halo geometry.
Among them, the best fits are obtained with  disk distributions extending over $\sim  200^\circ 
\times 5-6 ^\circ$ (gaussian shapes, 240$\mu$ and 1.25$\mu$ NIR/DIRBE maps or Robin 1-3 Gyr models).
The disk flux is around 1.7 $\times 10^{-3}~photons~cm^{-2}~s^{-1}$ leading to a bulge-to-disk
flux ratio of 0.4. This one  ranges from 0.25 (for the 1.25$\mu$ NIR/DIRBE map which presents a broad
bump emission in the Galactic Centre region)  to 0.7 (for less extended distributions like 
$150^\circ \times 5^\circ$ gaussian). 
This kind of disk emission morphology, widened in longitude, suggests
an old stellar population as the main Galactic positron source.\\
When allowing a halo contribution (Robin stellar halo model), the flux reaches a value of
$\sim 2.9 \times 10^{-3}~photons~cm^{-2}~s^{-1}$, with bulge-to- disk 
flux ratios evolving from 1.7 to 2.1 with
the same trend than above.

%It ranges from $\sim$ 1.7 for disk geometries to 2.9$\times 10^{-3}~photons~cm^{-2}~s^{-1}$ when including a halo component
%(Robin stellar halo model).
With our best fit determination, the disk is found to be symmetric with a west to east ratio of 1.1.
It is clear that the disk results are very sensitive to the assumed bulge morphology, with a strong coupling between
the parameters of both structures.
Fig.~\ref{fig:fig_ratio} illustrates that,
considering the error bars on the bulge position, it is impossible to claim a disk asymmetry.
The observed trend can be understood in terms of  
unavoidable cross-talk between  both emissions (bulge/disk) in the common region:
when the bulge is allowed to move,  the reconstructed disk flux is affected.

The INTEGRAL mission has significantly advanced positron astronomy. We begin to have a 
reliable view of the annihilation emission from the inner Galaxy, and can hope to possibly constrain
the extended emission.  It should bring some breakthrough in the quest for the mysterious origin of the 
positrons. At the same time, spectroscopy of the annihilation emission has yielded important insights 
into the conditions of the medium in which the positrons annihilate (Churazov et al., 2005, Jean et al., 2006).
Our work can be summarized as follows:

\begin{itemize}
\item 511 keV emission is significantly detected ($~70 \sigma$ in data space) towards
the galactic bulge region, and, at a lower level ($25 \sigma$) from the Galactic disk

\item There is no evidence for a point-like source in addition
to the diffuse emission, down to a typical flux limit of $4 \times 10^{-5}~photons~cm^{-2}~s^{-1}$ ($2\sigma$).

\item Among the variety of possible disk geometries, we favor extended ($\sim  200^\circ 
\times 5-6 ^\circ$) disk distributions, suggesting  
an old stellar population as the main Galactic positron source. The bulge-to-disk
flux ratio ranges from 0.25  to 0.7.\\
When allowing a halo contribution, the bulge-to- disk 
flux ratios evolves from 1.7 to 2.1.

\item For our reference model (2 gaussians bulge + Robin 1-2 Gyr disk),
we obtain  annihilation rates of $1.1\times 10^{43} s^{-1}$ in the bulge and $0.8\times 10^{43} s^{-1}$
in the disk.
 
\item The disk asymmetry reported in Weidenspointner et al. (2008) is not supported by our analysis.

\item The bulge emission appears almost spherically symmetric around
the Galactic Gentre with an extension of $12^\circ$. Its  centre seems slightly shifted towards negative longitudes at 
$l=-0.64\pm0.20^\circ$.

\end{itemize}

%\begin{acknowledgements}
\section*{Acknowledgments}  The \textit{INTEGRAL} SPI project has been completed under the responsibility
and leadership of CNES. We are grateful to ASI, CEA, CNES, DLR, ESA, INTA, NASA and OSTC for support.

%\end{acknowledgements}

%%%%%%%%%%%%%%%%%%%%%%%%%%%%%%%%%%%%%%%%%%%%%%%%%%%%
%%%%%%%%%% Figures 
%%%%%%%%%%%%%%%%%%%%%%%%%%%%%%%%%%%%%%%%%%%%%%%%%%%%

%%%%%%%%%%%%%%%%%%%%%%%%%%%%%%%%%%%%%%%%%%%%%%%%%%%%%%%%%
%%   TABLE
%%%%%%%%%%%%%%%%%%%%%%%%%%%%%%%%%%%%%%%%%%%%%%%%%%%%%%%%%

\begin{deluxetable}{cccccc}%1
\tablewidth{0pt}
\tablecaption{Basic empty-field dataset construction.} 

\tabletypesize{\scriptsize}
\tablehead{
\colhead{Inter annealing } 
&\colhead{Empty-field limits} 
&\colhead{Data total} 
&\colhead{Percentage of }
&\colhead{Empty-field exposures distribution}  \\

\colhead{period number }
&\colhead{First-last revolutions} 
&\colhead{exposure number} 
&\colhead{empty-field}   
&\colhead{$l < 0^\circ /$l$> 0^\circ /$b$< 0^\circ /$b$> 0^\circ$}  } 

\startdata
1        & ~44 - ~92    &  3328 &     12.6 &   192 /  226 /  ~~8 /  410 \\
2        & ~97 - 131    &  1816 &     ~8.4 &   ~58 /  ~95 /  101 /  ~52 \\

3$^a$    & 137 - 139    &   ~   &    ~     &   ~                        \\
         & 140 - 204    &  4334 &     ~5.2 &   128 /  ~97 /  129 /  ~96 \\

4$^b$    & 210 - 214    &  ~277 &     ~8.7 &   ~~0 /  ~24 /  ~24 /  ~~0 \\
         & 215 - 277    &  3730 &     13.0 &   218 /  268 /  181 /  305 \\

5        & 283 - 325    &  2767 &     18.8 &   199 /  321 /  ~72 /  448 \\
6        & 331 - 394    &  3550 &     10.6 &   138 /  240 /  186 /  192 \\
7        & 401 - 445    &  2668 &     15.2 &   296 /  110 /  ~~0 /  406 \\
8        & 452 - 505    &  3093 &     15.2 &   278 /  191 /  401 /  ~68 \\
9        & 512 - 564    &  3309 &     ~8.0 &   ~51 /  213 /  ~46 /  218 \\
10       & 571 - 640    &  4364 &     19.2 &   342 /  495 /  261 /  576 \\
11       & 647 - 713    &  3361 &     25.0 &   425 /  414 /  252 /  587 \\
12     & 720 - 775  &  2697 &     36.9 &   877 /  117 /  407 /  587 \\
\enddata
\tablecomments{
$^a$ detector 2 failure on Dec. 7, 2003 (between revolution 139 and 140) , $^b$ detector 17
failure on Jul. 17, 2004 (between revolution 214 and 215).
There is enough empty-field exposures to compute the detector pattern before (revolutions 210-214) 
and after (revolutions 215-277) the failure of detector 17. 
In the case of detector 2 failure, the detector pattern can be derived  
only after the revolution 140 (revolutions 140-204).}

\end{deluxetable}

\end{document}